\documentclass[12pt]{iopart}
\usepackage{iopams}
\usepackage{graphicx}
\usepackage{epsf}

\begin{document}

\sloppy

\jl{2}

\title[The role of inner-shell excitations in electron recombination
with multiply charged complex ions]
{\bf The role of inner-shell excitations in electron recombination
 with complex multicharged ions}

\author{S. Sahoo and G. F. Gribakin}

\address{Department of Applied Mathematics and Theoretical Physics, 
Queen's University, Belfast BT7~1NN, UK}



\begin{abstract}
We performed an overview of the inner shell excitation phenomena in the
study of electron recombination with multiply charged complex ions such
as Au$^{q+}$ (q=49-52) and Pb$^{53+}$. It
is found the the inner shell excitations play a significant role in the
low energy electron recombination. Taking into account these inner shell excitations
we have calculated the energy avaraged capture cross sections. The
dielectronic  rate coefficients are found to be in good agreement with the
experiment. We show that the contribution from the inner-shell enhances the recombinmation
rate by an order of magitude. We also made an attempt to identify the
resonaces observed by the experiment in Au$^{50+}$ and Pb$^{53+}$. A
prediction is made for the rate enhancement in Au$^{52+}$ in which we
found the configuration mixing between the doubly and the complex
multiply excited states. Analyzing the
statistics of eigenstate components we estimated the spreading width
is about 0.85 a.u. which defines the energy range within which strong
mixing takes place. 
\end{abstract}
\vspace{1cm}


\section{Introduction}
The electron recombination process has been the most
thorougly studied problem in the present era due to its practical
importance in different areas of modern physics. In particiular
dielectronic recombination (DR) process is one of the important
recombination process in high temperature plasmas. Accurate DR rate 
coeifficients are needed for successful modeling of astrophysical and 
laboratory plasmas. The current interest in study of the DR is the rate
enhancement in the recombination of low energy electron with multiply 
charged complex ions. This enhancement phenomena was first observed by
Muller et al. \cite{Mul:1991} in U$^{28+}$ ion. The interest in the rate
enhancement is largly due to the importance of understanding the
mechanism of recombination in particular those involving multiply
charged complex ions. Appart from this the potential application in
the production of antihydrogen by recombination of an antiproton with
positrons has provided a great deal of stimulus to study the detaild
mechanism of the DR process \cite{MM:1999}. The main feature in the rate enhancement is 
that at low energy, the measured recombination rate coefficients 
significantly exceed the theoretical predictions for radiative
recombination (RR). In the higher energies, measured rates is well
explained by RR theory together with contribution of DR. 

The complex ions  such as Au$^{25+}$
\cite{Hof:1998}, Au$^{50+}$ \cite{Uwira:1997} and Pb$^{53+}$
\cite{Baird:1995,Lind:2001} show a strongly enhanced recombination
rate. In  Au$^{25+}$ ion the
rate enhancement has been justified by Gribakin et. al.\cite{Au} and
Flambaum et. al.\cite{Fl:2000} using a statistical
approach. They showed that the electron capture is mediated by complex
multiply excited states rather than sinple dielectrronic resonances in
this system. In a recent study Gribakin et. al. \cite{Sahoo} have
explained the detailed mechanism responsible for low energy rate
enhancement in U$^{28+}$. In this system these authors reported that the
mechansim lies behind the fact that doubly excited configurations mix
with each other weakly and they do mix
with the complex multiply excited states quite comfortably. Most
importantly the excitations from inner shell hole plays a major
role in the low electron energies which was absent in the calculation of
Mitnik et al \cite{Mitnik:98} and is unable to produce
the results in agreement with the experiment in the low electron energies. For Pb$^{53+}$ ion,
Lindorth et al \cite{Lind:2001} performed a comphrenssive study to calculate
the resonances using relativistic perturbation theory. They found on
comparision with the experiment, the energy splitting in
Pb$^{53+}$(4$p_{1/2}$-4$s_{1/2}$) with an accuracy comparable to the
position of first few resonances and concluded that such an accuracy
provides a test of QED in many body systems. Although these authors
able to identify the the resonances with a very good precision, their
theoretical values for recombination rate  still needs an improvement in
order to be in good agreement with the measured data. This system is of
the current interest in the present study along with another similar system
i.e., Au$^{50+}$ since there is no theory for this particular system. In
order to achive a claer understanding we have also investigated the
nearest neighbouring ions such as Au$^{49+,51+,52+}$. In the present
calculation we have included the excitations from the inner shells and
found that the results are significantly improved with this
inclusion. The importance of the inner shell ecitation in the low energy
electron recombination in U$^{28+}$ has already been established
\cite{Sahoo}. Moreover, a recent study on bound doubly excited states
formation when slow highly charged ions (Ta$^q+$,q = 39-48) capture a single electron in
collsion with He by Schuch et. al.\cite{Sch:2000} reports that they observed a
strog characterstics M x-Ray emission without an M shell vecancy
initially present.In the present article we present the results of
recombination rates for the above mentioned ions taking into account the
inner shell vecancy. We compare these results with the available calculations that
do not take into account the inner shell excitations. A good agreement
has been achieved in comparison with the available experimental data as well.

\section{Many-electron excitations}\label{sec:mix}

Ref. \cite{Au,Fl:2000} suggest that electron recombination with complex
multiply charged ions is mediated by
complex {\em multiply-excited} states of the compound ions (target+e)
rather than ``simple'' dielectronic resonances. Electrons could be captured
in these states due to a strong configuration interaction. The ground
state of Au$^{q+}$ for $(q=49-51)$ is described by the
$1s^2\dots 3d^{10}4s^k (k=0,1,2)$ configuration and for
Au$^{q+}$$(q=52)$, it is described by $1s^2\dots 3d^{k}(k=9)$
configuration. Similarly for Pb$^{q^{'}}(q^{'}=53)$, the ground state
configuration is $1s^2\dots 3d^{10}4s^k (k=1)$. Figure 1 shows the spectrum
of relativistic single-particle orbitals of Au$^{q+}$$(q=50)$. The occupied orbitals
(below the Fermi level) are obtained in a relativistic Dirac-Fock calculation
of the Au$^{q+}$ ground state, and the excited state orbitals (above
the Fermi level) are obtained by solving the Dirac-Fock equation for an
electron in the potential of Au$^{25+}$ $1s^2\dots 4s^{1}$ core. Tabe 1
shows the ground state energy obtained from configuration interaction (CI)
calculation for different targets and the compound ions. 
\begin{figure}[h]
\epsfxsize=10cm
\centering\leavevmode\epsfbox{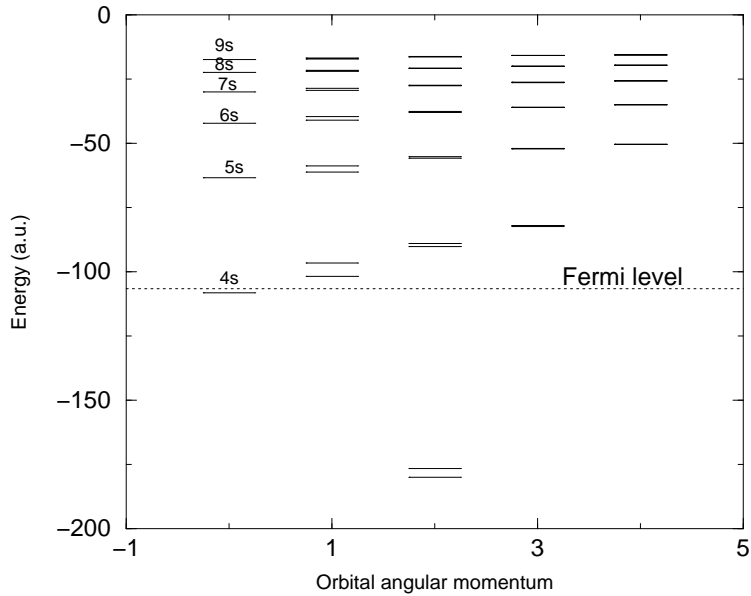}
\vspace{8pt}
\caption{Energies of occupied and vacant single-particle
orbitals of Au$^{50+}$ obtained in a Dirac-Fock calculation.}
\label{fig:orb}
\end{figure}
\begin{table}
\caption{Ground state energy and Ionization threshold (I) of
different ions.}
\label{tab:orb}
\begin{tabular}{cccc}
\multicolumn{4}{c}{Ground state Energy (a.u.)}\\
$q $ & Au$^{q+}$ & Au$^{(q-1)+}$ & I(a.u.)\\
\hline
49 & 17287.64 & 17387.22 & 99.58 \\ 
50 & 17181.91 & 17287.56 &105.65 \\
51 & 17073.75 & 17181.90 &108.15 \\
52 & 16894.25 & 17073.60 &179.35\\
\hline
 & Pb$^{q+}$ & Pb$^{(q-1)+}$ & I(a.u.) \\
\hline
53 & 18742.07 & 18860.26 & 118.19 \\
\end{tabular}
\end{table}
 The
difference between the total energies of the target ions and their
respective compound
ions gives us the ionization threshold which is shown in table I. Since
we are interested near the ionization threshold, we
consider 3p orbitals are inactive and allowed the orbitals from 3d
onwards to take active part in the process. We distributed these
active electron among 61 relativistic orbitals from $3d_{3/2}$ to $9g_{7/2}$.We construct the
excitation spectrum of Au$^{(q-1)+}$ and Pb$^{(q^{'}-1)+}$ by calculating the mean energies $E_i$, and number of many electron states $N_i$:
\begin{equation}\label{eq:E}
E_{i}= E_{core}+\sum_a{\epsilon_a n_a}+\sum_{a<b}\frac{n_a(N_b-\delta_{ab})}{1+\delta_{ab}}U_{ab} ,
\end{equation}
\begin {equation}\label{eq:N}
N_i=\prod_a\frac{g_a!}{n_a! (g_a-n_a)!},
\end {equation}
where $n_a$ are the orbital occupation numbers of the relativistic orbitals in a given configuration and $\sum_a{n_a}=n$. $\epsilon_a=<a|H_{core}|a>$ is the single-particle energy of the orbital $a$ in the field of the core, $g_a=2j_a+1$, and $U_{ab}$ are the avarage Coulomb matrix elements for the electrons in orbitals $a$ and $b$ ( direct minus exchange):
\begin{equation}\label{eq:U}
U_{ab}=\frac{g_a}{g_a-\delta_{ab}}\left[R_{abab}^{(0)}-\sum_{\lambda}\delta_p R_{abba}^{(\lambda)}\left\{ {j_a \atop \frac{1}{2}}{\j_b \atop-\frac{1}{2}}{\lambda \atop \ 0}\right\}^{2}\right]
\end{equation}
$R_{abba}^{(\lambda)}$ is the two-body radial Coulomb integral of $\lambda$ multipole, and $\delta_p=1$ when $l_a +l_b +\lambda$ is even and 0 otherwise.
Using the single-particle orbitals we have generated many-electron
configurations, evaluated their energies and estimated the energy density of
multiply-excited states \cite{Au}. 
\begin{figure}[h]
\epsfxsize=10cm
\centering\leavevmode\epsfbox{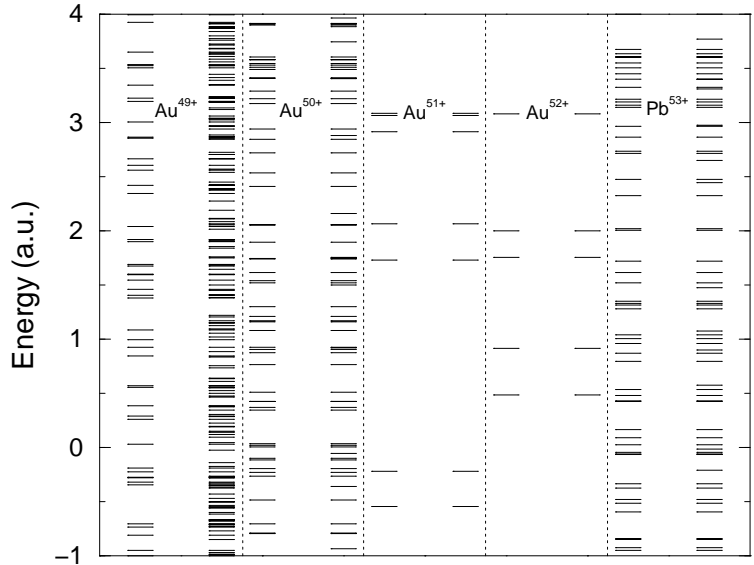}
\vspace{8pt}
\caption{ Position of the dielectronic and multiply excited states of
ions with respect to their ionization threshold (see table I).}
\end{figure}

Figure 2 shows the doubly excited
states as well as multiply excites states near the ionization threshold
of the ions. We plotted all the configurations having energies between
-1 and 4 a.u.. In Au$^{49+}$, there is only one doulby excited state
near the ionization threshold although a significant number of multiply
excited states are found nearby this particular dielectronic state. It
is also found that the multiply excited states are not populated through
this state and hence rate enhancement is not expected, which has been
explained by the experiment\cite{Uwira:1997}. In case of Au$^{51+}$, we did not
find any dielectronic states near its ionization threshold. However,
Au$^{50+}$and Pb$^{53+}$, shows a number of dielectronic states near its
ionization threshold and these contribute largly to the rate
enhancement. Au$^{52+}$ shows there are a few dielectronic states above
the ionization threshold. It may be mentioned that one may expect rate
enhancement in this ion because these dielectronic states shows a
substantial mixing with each other. The detailed study of the
nature dielectronic states near the ionization threshold of all the ions
is disccussed later. 
\section{Recombination}\label{sec:rec}
\subsection{Theory}
For low-energy electrons the contribution of the autoionising
states (resonances) to the recombination cross section is given by 
(see, e.g., Ref. \cite{Landau})
\begin{equation}\label{eq:sigma_res}
\sigma _r=\frac{\pi }{k^2}\sum _\nu \frac{2J+1}{2(2J_i+1)}\,
\frac{\Gamma _\nu ^{(r)}\Gamma _\nu ^{(a)}}{(\varepsilon -\varepsilon _\nu )^2
+\Gamma _\nu ^2/4},
\end{equation}
where $\varepsilon =k^2/2$ is the electron energy, $J_i$ is the angular
momentum of the initial (ground) target state, $J$ are the angular momenta
of the resonances, $\varepsilon _\nu =E_\nu -I$ is the position of the
$\nu $th resonance relative to the ionization threshold of the compound
(final-state) ion, and $\Gamma _\nu ^{(a)}$, $\Gamma _\nu ^{(r)}$,
and $\Gamma _\nu =\Gamma _\nu ^{(r)}+\Gamma _\nu ^{(a)}$ are its
autoionisation, radiative, and total widths, respectively
\cite{comment4}. When the resonance spectrum is dense, $\sigma _r$
can be averaged over an energy interval $\Delta \varepsilon$ which 
contains many resonances,
$D\ll \Delta \varepsilon \ll \varepsilon $, yielding
\begin{equation}\label{eq:sigres_av}
\bar \sigma _r=\frac{2\pi ^2}{k^2}\sum _{J^\pi }
\frac{2J +1}{2(2J_i+1)D}\left\langle \frac{\Gamma _\nu ^{(r)}
\Gamma _\nu ^{(a)}}{\Gamma _\nu ^{(r)}+\Gamma _\nu ^{(a)}} \right\rangle ,
\end{equation}
where $\langle \dots \rangle $ means averaging. If the fluorescence yield,
$\omega _f\equiv \Gamma _\nu ^{(r)}/(\Gamma _\nu ^{(r)}+\Gamma _\nu ^{(a)})$,
fluctuates weakly from resonance to resonance (see below), one can
write $\bar \sigma _r=\bar \sigma _c\omega _f$, where
\begin{equation}\label{eq:sig_cap}
\bar \sigma _c=\frac{\pi ^2}{k^2}\sum _{J^\pi } \frac{(2J +1)
\Gamma ^{(a)}}{(2J_i+1)D}
\end{equation}
is the energy-averaged capture cross section, and $\Gamma ^{(a)}$ is the
average autoionisation width.

In a situation when there is a strong configuration mixing between the
dielectronic doorway states and multiply excited states, the capture
cross sections can be obtained as a  sum over the single-electron excited states
$\alpha ,~\beta $ and hole states $\gamma $, as well as the partial waves
$lj$ of the continuous-spectrum electron $\varepsilon $. As a result, we have
\begin{eqnarray}\label{eq:sig_cap2}
\bar \sigma _c&=&\frac{\pi ^2}{k^2}\sum _{\alpha \beta \gamma ,lj}
\frac{\Gamma _{\rm spr}}{(\varepsilon -\varepsilon _\alpha -
\varepsilon _\beta +\varepsilon _\gamma )^2 +\Gamma _{\rm spr}^2/4}
\sum _\lambda \frac{\langle \alpha ,\beta \| V_\lambda \| \gamma ,
\varepsilon lj \rangle }{2\lambda +1}\nonumber \\
&\times &\Biggl[ \langle \alpha ,\beta \| V _\lambda \| \gamma ,
\varepsilon lj\rangle -(2\lambda +1) \sum _{\lambda '}
(-1)^{\lambda +\lambda '+1}\left\{ {\lambda \atop \lambda '}{j_\alpha 
\atop j_\beta }{j \atop j_\gamma }\right\} \langle \alpha ,\beta \| V
_{\lambda '}\| \varepsilon lj,\gamma \rangle \Biggr] ,
\end{eqnarray}
where $\varepsilon _\alpha $, $\varepsilon _\beta $ and $\varepsilon _\gamma $
are the orbital energies, the two terms in square brackets represent the
direct and exchange contributions, and
$\langle \alpha ,\beta \| V_\lambda \| \gamma ,
\varepsilon lj \rangle $ is the reduced Coulomb matrix element.

It is assumed that the energies of dielectronic doorway states relative
to the threshold is given by $\varepsilon _\alpha $ + $\varepsilon
_\beta $ - $\varepsilon _\gamma$. 
A more accurate value can be obtained by using mean field energies
 (configuration energies) of
doorway configurations in Eq.(10).  Therefore we used the configuration
energy in the calculation of capture cross sections.    

Equation (\ref{eq:sig_cap2}) is directly applicable to targets with
closed-shell ground states. If the target ground state contains partially
occupied orbitals, a factor
\begin{equation}\label{eq:occup}
\frac{n_\gamma }{2j_\gamma +1}\left(1-\frac{n_\alpha }{2j_\alpha 
+1}\right)\left(1-\frac{n_\beta }{2j_\beta  +1}\right),
\end{equation}
where $n_\alpha $, $n_\beta $, and $n_\gamma $ are the orbital occupation
numbers in the ground state $\Phi _i$, must be introduced on the right-hand
side of Eq. (\ref{eq:sig_cap2}). Steps similar to those that lead to
Eq. (\ref{eq:sig_cap2}) were used to obtain mean-squared matrix elements of
operators between chaotic many-body states \cite{Ce,Flambaum:93}.

The chaotic nature of the multiply-excited states $\Psi _\nu $ can also be
employed to estimate their radiative widths $\Gamma _\nu ^{(r)}$. 
Electron-photon interaction is described by a single-particle dipole operator
$\hat d$. Any excited electron in $\Psi _\nu $ may emit a photon, thus leading
to radiative stabilization of this state. The total photo-emission rate
$\Gamma _\nu ^{(r)}$ can be estimated as a weighted sum of the single-particle
rates,
\begin{equation}\label{eq:Gamma_r}
\Gamma _\nu ^{(r)}\simeq \sum _{\alpha ,\beta }
\frac{4\omega _{\beta \alpha }^3} {3c^3}
|\langle \alpha \|\hat d\|\beta \rangle |^2
\left\langle \frac{n_\beta }{2j_\beta +1}\left( 1-\frac{n_\alpha }
{2j_\alpha +1}\right) \right\rangle _\nu ,
\end{equation}
where $\omega _{\beta \alpha }=\varepsilon _\beta -\varepsilon _\alpha >0$,
$\langle \alpha \|\hat d\|\beta \rangle $ is the reduced dipole operator
between the orbitals $\alpha $ and $\beta $, and
$\langle \dots \rangle _\nu $ is the
mean occupation number factor. Since $\Psi _\nu $ have large numbers
of principal components $N$, their radiative widths display small $1/\sqrt{N}$
fluctuations. This can also be seen if one recalls that a chaotic
multiply-excited state is coupled by photo-emission to many lower-lying states,
and the total radiative width is the sum of a large number of (strongly
fluctuating) partial widths. A similar effect is known in compound nucleus
resonances in low-energy neutron scattering \cite{Bohr:69}.

There is a certain similarity between Eqs. (\ref{eq:sig_cap2}) and
(\ref{eq:Gamma_r}) and those for autoionisation and radiative rates obtained
in a so-called configuration-average approximation \cite{Griffin:85}.
In both cases the answers involve squares or products of two-body Coulomb
matrix elements [see the direct and exchange terms in Eq. (\ref{eq:sig_cap2})],
or single-particle dipole amplitudes [Eq. (\ref{eq:Gamma_r})]. However, there
are a number of important differences between the present results and
the configuration-average approximation. The latter considers dielectronic
recombination and introduces averaging over configurations as a means of
simplifying the calculation. The DR cross section is averaged over an
arbitrary energy interval $\Delta \varepsilon $, and only the configurations
within this energy range contribute to the average. Effects of configuration
mixing as well as level mixing within a configuration are neglected.

It is important to compare the radiative and autoionisation widths of
chaotic multiply-excited states. Equation (\ref{eq:Gamma_r}) shows that
$\Gamma ^{(r)}$ is comparable to the single-particle radiative widths.
On the other hand, the autoionisation width $\Gamma ^{(a)}$, is suppressed by a factor
$\left| C_k^{(\nu )}\right| ^2\sim N^{-1}$ relative to that of a typical
dielectronic resonance.  Therefore, in systems
with dense spectra of chaotic multiply-excited states the autoionisation
widths are small. Physically this happens because the coupling strength
of a two-electron doorway state to the continuum is shared between many
complicated multiply-excited eigenstates.
As a result, the radiative width may dominate the total width of the
resonances, $\Gamma ^{(r)}\gg \Gamma ^{(a)}$, making their
fluorescence yield close to unity. 

The resonance recombination cross section should be compared with the 
direct radiative recombination cross section
\begin{equation}\label{eq:sigmad}
\sigma _d= \frac{32\pi }{3\sqrt{3}c^3}\,\frac{Z_i^2}
{k^2} \ln \left( \frac{Z_i}{n_0k}\right) ,
\end{equation}
obtained from the Kramers formula by summing over the principal
quantum number of the final state \cite{Au}. Note that the direct 
and energy-averaged resonance recombination cross sections of
Eqs. (\ref{eq:sigmad})
and  (\ref{eq:sigres_av}) have similar energy dependences.

\subsection{Numerical results}\label{sec:num}
 \underline{Au$^{49+}$}\\

 We present the RR and DR rate coefficients for Au$^{49+}$ in figure
 3 in the energy range between 0 to 100 eV. We also included in the
 same figure the dielectronic configurations producing the resonances
 and are by solid vertical lines. The position of peaks in the DR rate
 corresponds to each of the doubly excited states. It is
 evident from the figure that there is no strong dielectonic
 configurations near
\begin{figure}[h]
\epsfxsize=10cm
\centering\leavevmode\epsfbox{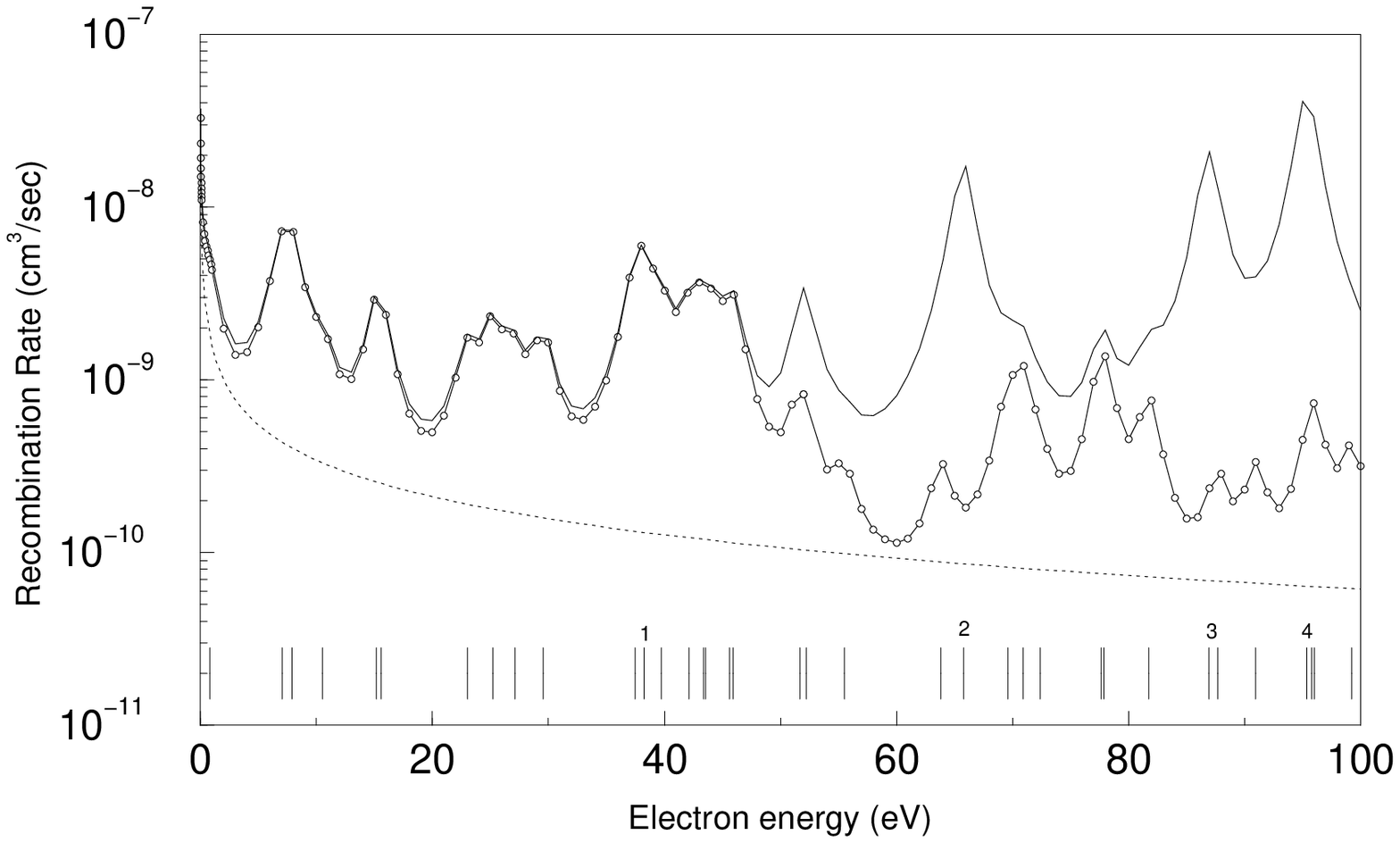}
\vspace{8pt}
\caption{Recombination rate in Au$^{49+}$. dotted line: RR rate,
solid line: DR rate with a hole in inner shell (3d), open
circles connected by solid line: DR rate without inner shell excitation
and solid vertical lines numbered 1-4: position of dielectronic states
involving inner shell
hole.1: 3$d^{4}_{3/2}$3$d^{5}_{5/2}$4$s^{2}_{1/2}$4$d^{1}_{3/2}$4$f^{1}_{5/2}$,
2:
 3$d^{4}_{3/2}$3$d^{5}_{5/2}$4$s^{2}_{1/2}$4$d^{1}_{3/2}$4$f^{1}_{7/2}$,
3: 3$d^{4}_{3/2}$3$d^{5}_{5/2}$4$s^{2}_{1/2}$4$d^{1}_{5/2}$4$f^{1}_{5/2}$,
4: 3$d^{4}_{3/2}$3$d^{5}_{5/2}$4$s^{2}_{1/2}$4$d^{1}_{3/2}$4$f^{1}_{7/2}$.}
\end{figure}

 the ionization threshold and hence the experiment does not show an
 enhancement in the rate coefficients as well as  resonance structure in
 the low energies. We found the only isolated doubly excited state near the threshold 
 involves
 4s excitation is not particularly strong. However, at higher electron
 energies, quite a number of resonaces are observed. Of particular
 importance are the dielectronic states those produce larger peaks. We
 identified them as the states which involve the inner shell hole. These
 states are indicated with the numbers (1,2,3 etc) in the figure. One of
 such states found to appear around 40 eV is
 3$d^{4}_{3/2}$3$d^{5}_{5/2}$4$s^{2}_{1/2}$4$d^{1}_{3/2}$4$f^{1}_{5/2}$
 due to which a large peak is observed. We found in the present ion,
 these type of configurations give most contribution to the
 recombination rate. Identifying these type of configuration we plotted
 them with respect to the ionization threshold in Fig.8. It appears
 that configurations are far above the ionization threshold i.e above 40
 eV. To get more evidence if they produce any low energy resonance we
 performed a configuration calculation taking only these type of
 configurations into account. We obtained the excited spectrum for all
 total angular momentum J
 resulted from diagonilazation of configuration Hamiltonian. These are
 plotted  as a
 function of eigen energies ranging from -4 to 4 a.u. in Fig.9  and
 -2 to 2 eV in Fig.10. We found some resonances but  they are not
 particularly strong in nature. It is also interesting to note that we
 performed a calculation without including the inner shell
 excitations. The  results of rate coefficients are shown by circles connected by solid
 lines. This shows that the inner shell excitations in this ion plays a
 crucial role in the high electron energies and at low electron energies
 a negligible effect is found. This because the configurations are
 positioned in the higer electron energies (fig. 8).\\ 
 
\underline{Au$^{51+}$}\\
  
For the case of Au$^{51+}$ we found practically no dielectronic states
near its ionization threshold. As mentioned earlier large contributions to
the capture cross sections comes from the dielectronic states involving
the inner shell hole (3d) and 4d and 4f excited orbitals, these are found to be
in either side of the ionization threshold. This may be seen in Fig.4 wehere we have shown the
recombination rate (fig. 8). Figure 9 and figure 10  also indicate that there
is not even a single resonance around the threshold. 
\begin{figure}[h]
\epsfxsize=10cm
\centering\leavevmode\epsfbox{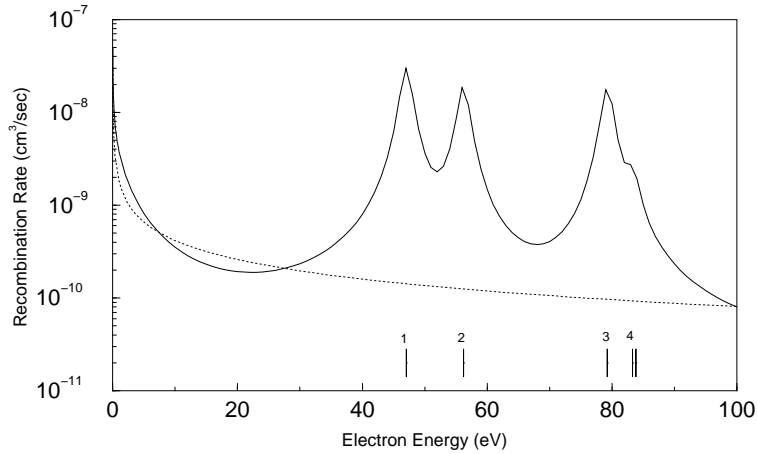}
\caption{Recombination rate in Au$^{51+}$. dotted line: RR rate,
solid line: DR rate. Vetical lines with numbers 1-4 indicate the position of
configurations involving inner shell hole. 1:
 3$d^{3}_{3/2}$3$d^{6}_{5/2}$4$d^{1}_{3/2}$4$f^{1}_{5/2}$, 2:
3$d^{3}_{3/2}$3$d^{6}_{5/2}$4$d^{1}_{3/2}$4$f^{1}_{7/2}$,
3: 3$d^{3}_{3/2}$3$d^{6}_{5/2}$4$d^{1}_{5/2}$4$f^{1}_{5/2}$, 4: 3$d^{3}_{3/2}$3$d^{6}_{5/2}$4$d^{1}_{5/2}$4$f^{1}_{7/2}$.}
\label{fig:orb}
\end{figure}
At low energies the
DR and RR rates are nearly equal. At higher energies the enhancement in
DR rate is found due to the presence of configurations involving the
inner shell hole. The position of each peaks in the DR rate coefficients
corresponds to the position of dilectronic configurations and are
indicated by vertical solid
lines. \\

\underline{Au$^{50+}$}\\    

Now we present the intersting results for Au$^{50+}$ in figure 5. We found there is a low energy DR rate enhancement which is in
agreement with experiment. 
\begin{figure}[h]
\epsfxsize=10cm
\centering\leavevmode\epsfbox{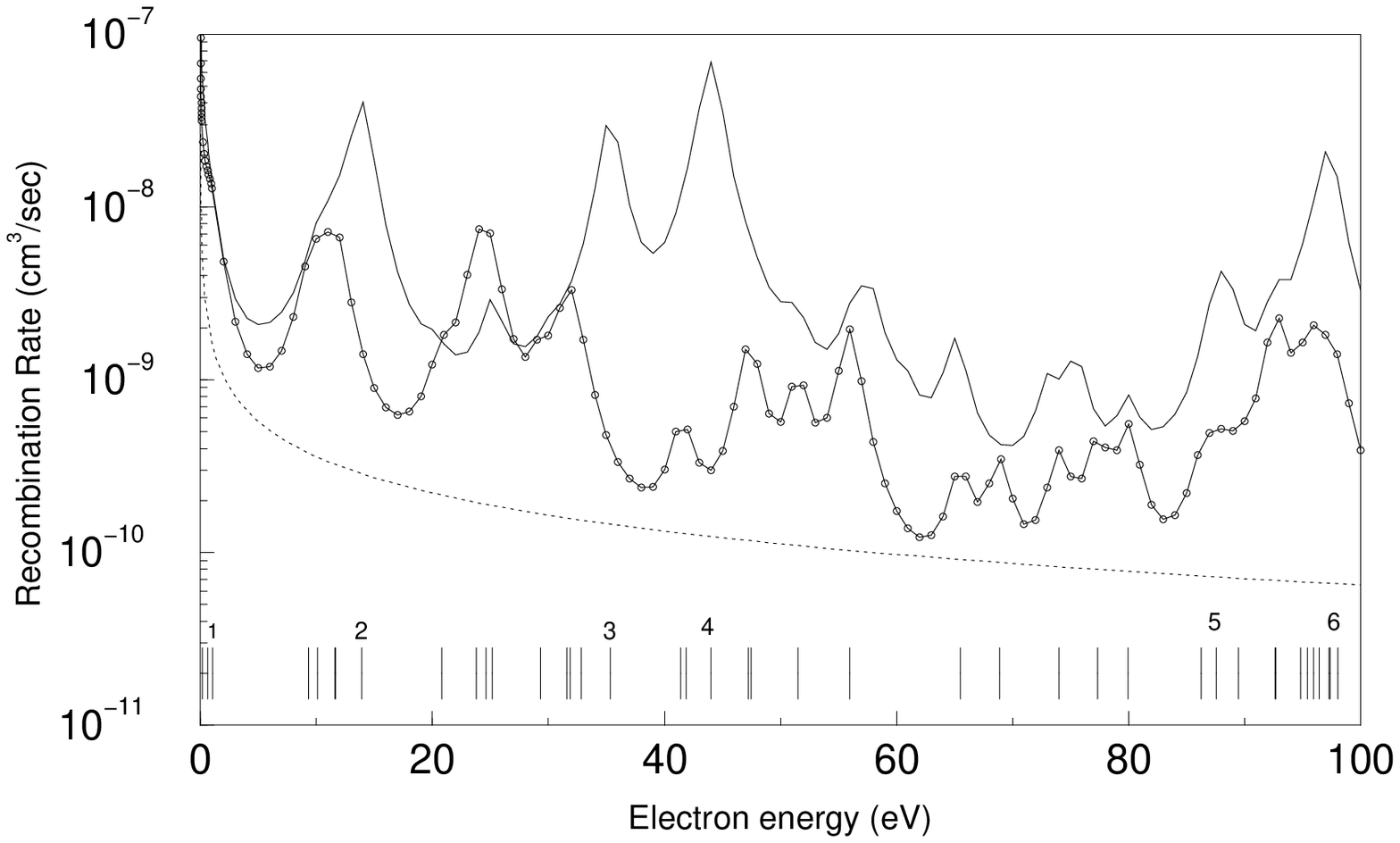}
\caption{Recombination rate in Au$^{50+}$. dotted line: RR rate,
solid line: DR rate with a hole in inner shell (3d), open
circles connected by solid line: DR rate without inner shell excitation
and solid vertical lines numbered 1 -6 : position of dielectronic states
involving the inner shell hole.1: 3$d^{4}_{3/2}$3$d^{5}_{5/2}$4$s^{1}_{1/2}$4$d^{1}_{3/2}$4$f^{1}_{5/2}$,
2:
3$d^{4}_{3/2}$3$d^{5}_{5/2}$4$s^{1}_{1/2}$4$d^{1}_{3/2}$4$f^{1}_{7/2}$,
3: 3$d^{4}_{3/2}$3$d^{5}_{5/2}$4$s^{1}_{1/2}$4$d^{1}_{5/2}$4$f^{1}_{5/2}$,
4: 3$d^{4}_{3/2}$3$d^{5}_{5/2}$4$s^{1}_{1/2}$4$d^{1}_{3/2}$4$f^{1}_{7/2}$,
5:
 3$d^{3}_{3/2}$3$d^{6}_{5/2}$4$s^{1}_{1/2}$4$d^{1}_{3/2}$4$f^{1}_{5/2}$,
6: 3$d^{3}_{3/2}$3$d^{6}_{5/2}$4$s^{1}_{1/2}$4$d^{1}_{3/2}$4$f^{1}_{7/2}$.}
\end{figure}

Also we show a lots of structures and huge
peaks in higher electron energies. We identified these peaks and found
they correspond to the doubly excited states involving inner shell
excitation. The huge enhancement is found around the positions of
configuration of type 3d4d4f. It will be interesting to look the Fig. 8 
where one can see that the configurations of above nature are close to the ionization threshold
and one of such state i.e.,
3d$^{4}_{3/2}$3d$^{5}_{5/2}$4s$^{1}_{1/2}$4d$^{1}_{3/2}$4f$^{1}_{5/2}$
is right at 0.01 a.u. above threshold. Due to its presence it 
is expectecd that one will obtain higher recombination rate which we
justified here. 
The experiment in the low energy found huge enhancement and around 10 
isolated resonances of which 3 resonances are strong. Our configuration 
interaction calculation shows that these states produce three resonances 
as shown in Fig. 10 between 0 and 1 eV of electron energy. 
The hole spectrum is shown in Fig 9. It is also interesting that when 
we switched off the inner shell excitation in our calculation there 
is significant  change in the
magnitude of DR rates throughout the energy range considerd. This
confirms the importance of inner shell excitations in electron
recombination with complex multiply excited states.\\

\underline{Pb$^{53+}$}\\ 

The schenerio discussed for Au$^{50+}$ is
even more interesting in the case of Pb$^{53+}$. The recombination rate
for this ion is presented in the figure 6. We found that there is huge
DR rate enhancement over RR background particularly in the low energy
region. The huge peaks in the recombination rate are found to be due to
the doubly excited states involving innershell (3d) excitations as
indicated earlier. If one looks at the figure 8, one can see that
just like Au$^{50+}$, in this system one such state
(3d$^{4}_{3/2}$3d$^{5}_{5/2}$4s$^{1}_{1/2}$4d$^{1}_{5/2}$4f$^{1}_{5/2}$) is positioned at
0.03 a.u. above ionization threshold and the rest configuration of this
kind are above and below the
ionization threshold. 
\begin{figure}[h]
\epsfxsize=10cm
\centering\leavevmode\epsfbox{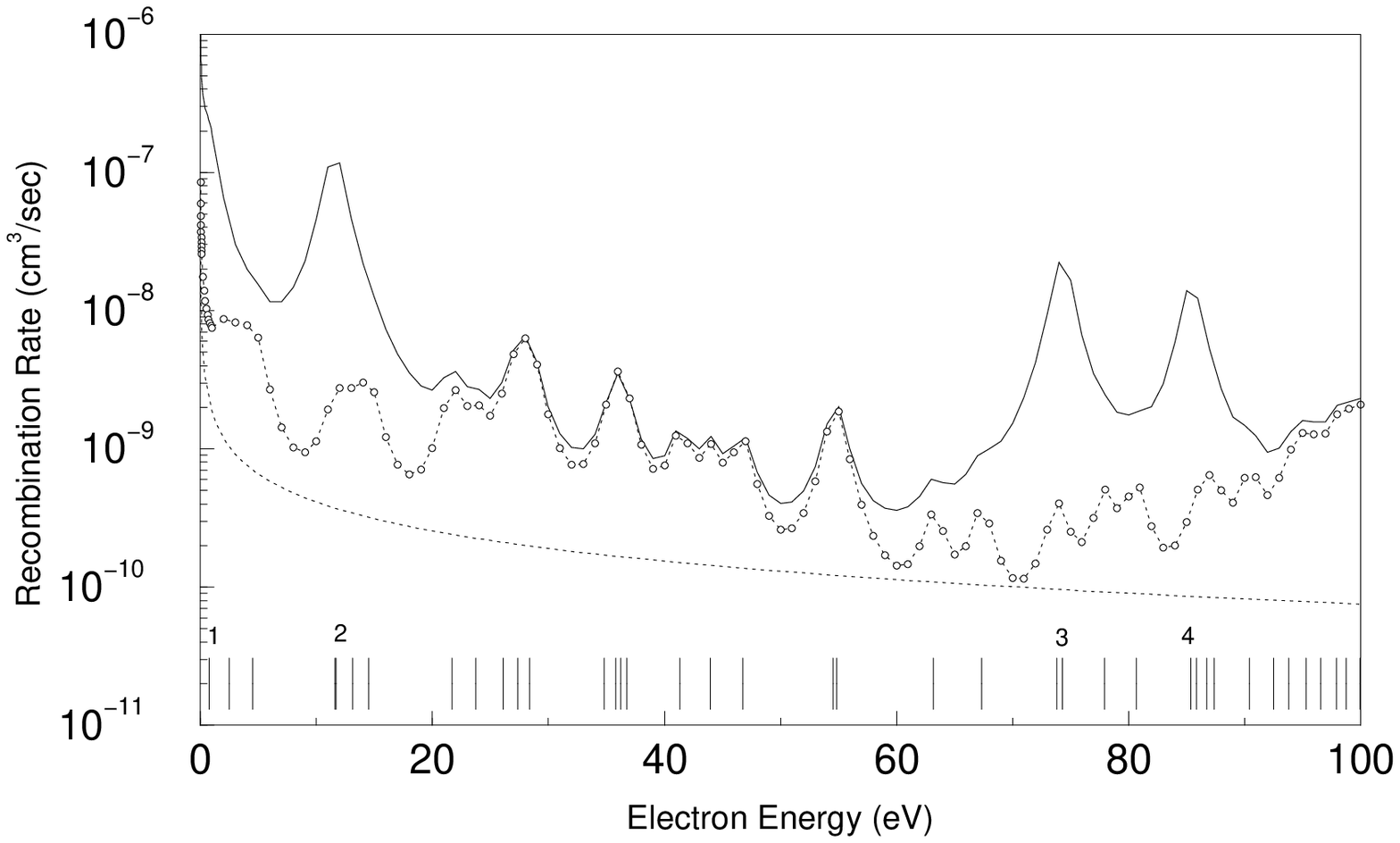}
\caption{Recombination rate in Pb$^{53+}$. dotted line: RR rate,
solid line: DR rate with a hole in inner shell (3d), open
circles connected by dotted line: DR rate without inner shell excitation
and solid vertical lines numberd 1-4: position of dielectronic states
involving innser shell hole.1: 3$d^{4}_{3/2}$3$d^{5}_{5/2}$4$d^{1}_{5/2}$4$f^{1}_{5/2}$, 2:
3$d^{4}_{3/2}$3$d^{5}_{5/2}$4$d^{1}_{5/2}$4$f^{1}_{7/2}$,
3:3$d^{3}_{3/2}$3$d^{6}_{5/2}$4$d^{1}_{3/2}$4$f^{1}_{5/2}$, 4: 3$d^{3}_{3/2}$3$d^{6}_{5/2}$4$d^{1}_{3/2}$4$f^{1}_{7/2}$.}

\end{figure}

This configuration at threshold
produces huge contributions to the recombination rate in the low energy and hence the
enhanced rate as found in the experiment. Another interesting phenomena
is that like Au$^{Au50}$, we also ignore the excitations in our
calculation and the results of which is indicated by small circles connected
by solid line. This line lies atleast factor of 10 below the results
that includes the inner shell hole
especially in the low energy region. It clearly signifies that one can
not ignore the inner shell excitations in electron recombination with
multiply charged ions. It may
be mentioned that the elaborate calculation \cite{Lind:2001} finds the
resonance peaks with a greater accuracy. However, there is still a
discrepency in magnitude
in comparison to the experimental data. They have taken all the rydberg
series into account and their ground state configuration does not
include any inner shell hole. To have close look at their results we compared
the present results in the low energy region between 0 to 0.1 eV electron
energy with the experimeental data \cite{Lind:2001} as shown in figure
7. 
\begin{figure}[h]
\epsfxsize=10cm
\centering\leavevmode\epsfbox{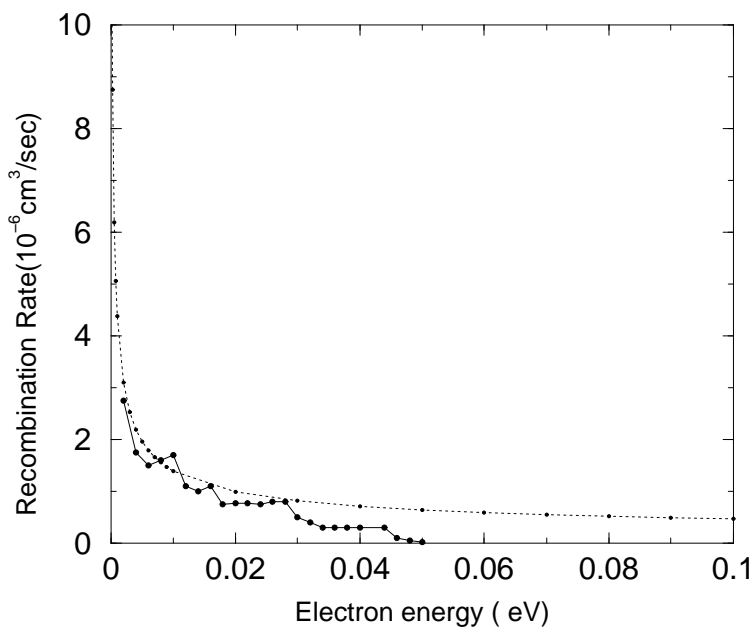}
\caption{DR rate of Pb$^{53+}$. Squares connexted by dotted line: Experimenatl
data [6], circles connected by solid line : Present calculation.} 
\end{figure}

It may be seen that the presnt results are found to be in
resonably good agreement with the experiment. \\

\begin{figure}[h]
\epsfxsize=10cm
\centering\leavevmode\epsfbox{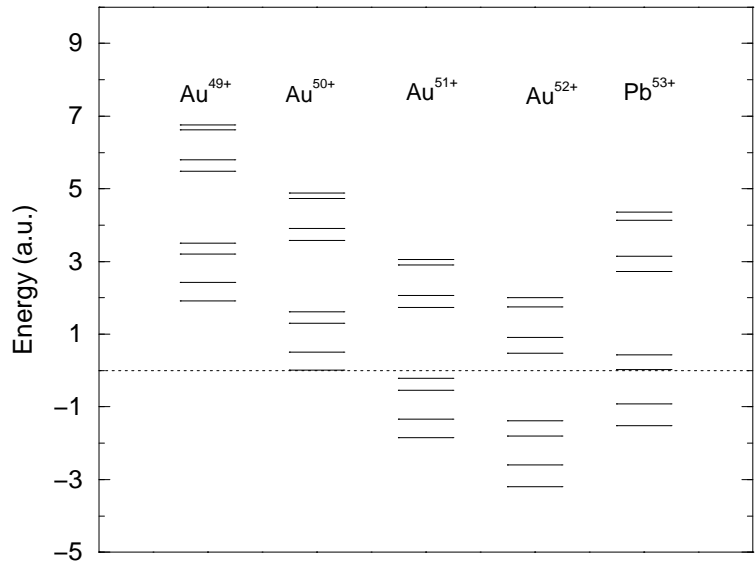}
\caption{Configurations  near the ionization threshold ions involving
only 3d holes. In each of the cases  top four horizontal lines indicate
3$d^{3}_{3/2}$ excitations with 4$d$4$4$ and the bottom four lines
represnt the position of the configurations involving 3$d^{5}_{5/2}$
excitations with 4$d$4$f$.}
\end{figure}

\begin{figure}[h]
\epsfxsize=10cm
\centering\leavevmode\epsfbox{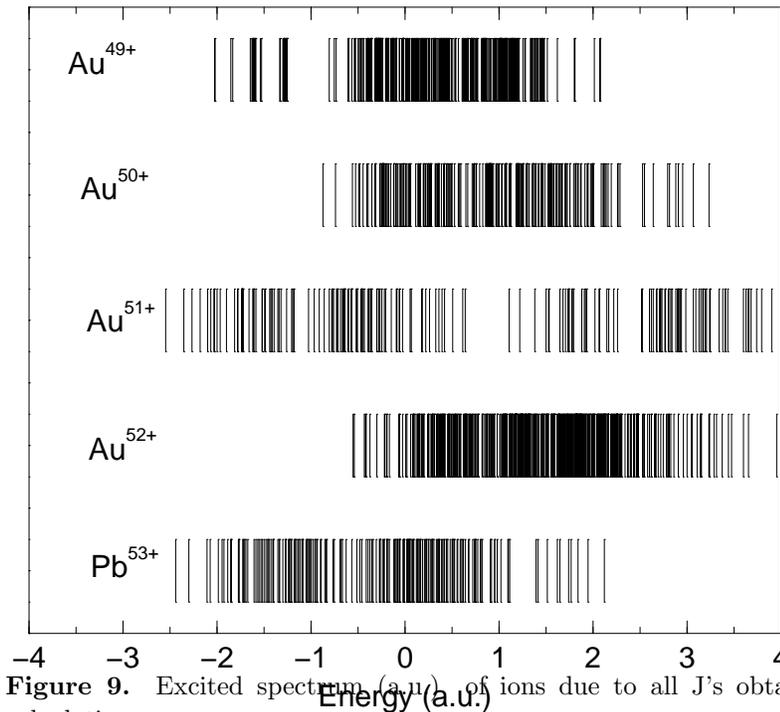}
\caption {Excited spectrum (a.u.) of ions due to all J's obtained from the CI
calculation.}
\end{figure}
\begin{figure}[h]
\epsfxsize=9cm
\centering\leavevmode\epsfbox{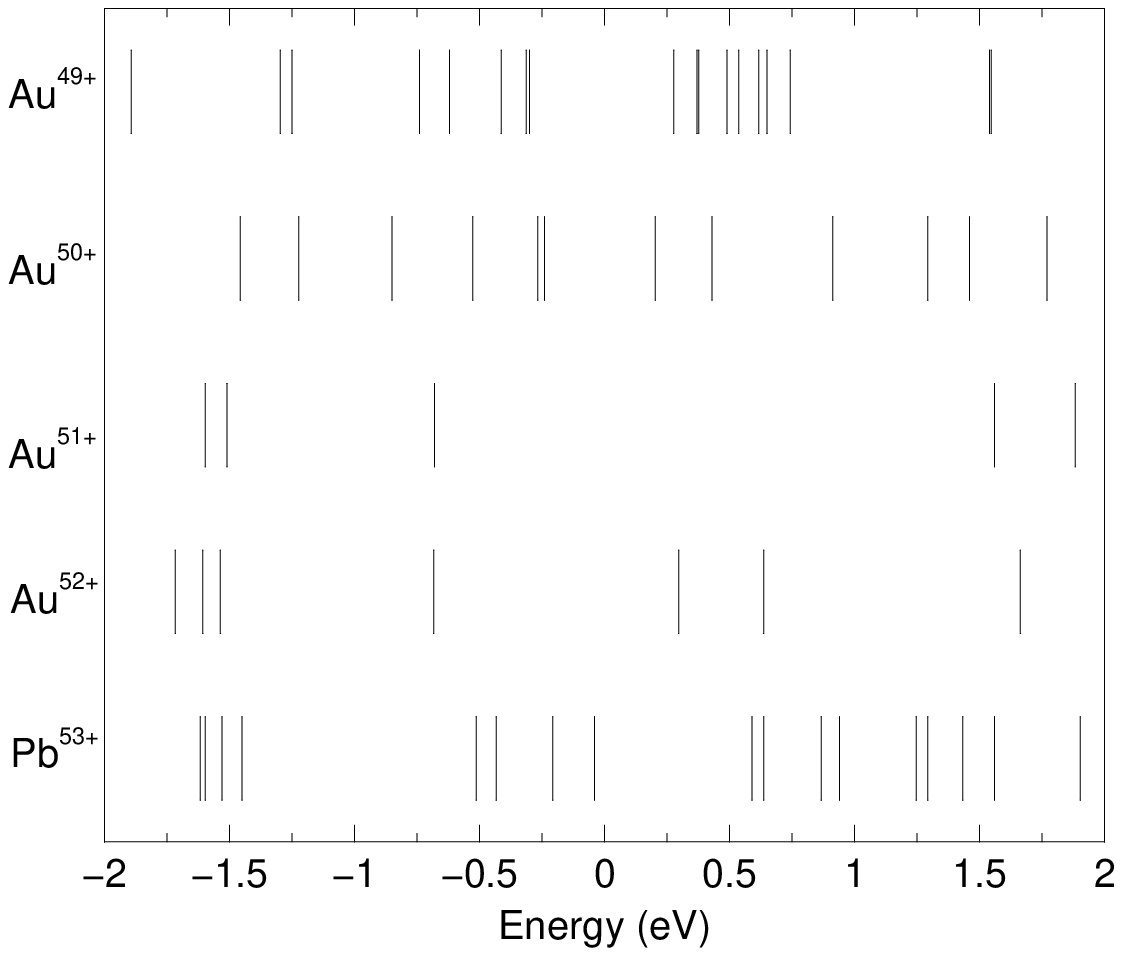}
\caption{Same as Figure 9 but in eV.}
\end{figure}

\underline{Au$^{52+}$}\\
         
We present the new results for recombination rate of Au$^{52+}$. Figure
11 shows the recombination rates as a function of electron energy. We
also display the position of the dielectronic states by vertical
lines those corresponds to each of peaks. In figure 2 we plot the dielectronic as well as multiply excited
states in the energy range of -1 to 4 a.u. with respect to the ionization
threshold (table 1) of this ion. It may be seen that there are not many doubly
excite states so as multiply excited states near the ionization threshold. However, similar to the
other ions discussed above we found the dielectronic configutrations involving an inner shell
excitations of type
3$d^{3}_{3/2}$3$d^{5}_{5/2}$4$d^{1}_{3/2}$4$f^{1}_{5/2}$ or 3$d^{4}_{3/2}$3$d^{4}_{5/2}$4$d^{1}_{3/2}$4$f^{1}_{5/2}$
contributes most to the recombination rates. Their positions are far
above and below the ionization threshold (Fig.8) and the closest being
around 0.5 a.u. ($\sim$ 13.6 eV) above the ionization threshold. Their
contribution 
can be seen in the recombination rate indicated by dashed lines
connected by open circles. Please note that this calculation uses an
arbitrary spreading width $\sim$ 0.75 a.u.. Taking these cofigurations
involving inner shell hole we performed a configuration calculation. Our
calculation shows that these dielectronic states mix with each other
quite comfortablly. To study the mixing we plotted the weights of these
dielectronic states in Fig. 12. This figure
indicates that there is a regular mixing between these states since the
weights has gone down significantly from 1. This figure also indicates
that these states spread largly on the energy scale. To estimate the spreading
width we calculated the mean-squared components and plotted them as a
function of basis state energy. The statistics of the meansquared
components is well approximated by Breit-Wigner (B-W) formula
(Fig. 13). From the BW fitting we estimated the spreading width is 0.85
a.u. which defines the energy range within which strong mixing takes
place. In the case of Au$^{25+}$, it was about 0.5 a.u. \cite{Au,JP}. 
Using this value of the spreading width

\begin{figure}[h]
\epsfxsize=9cm
\centering\leavevmode\epsfbox{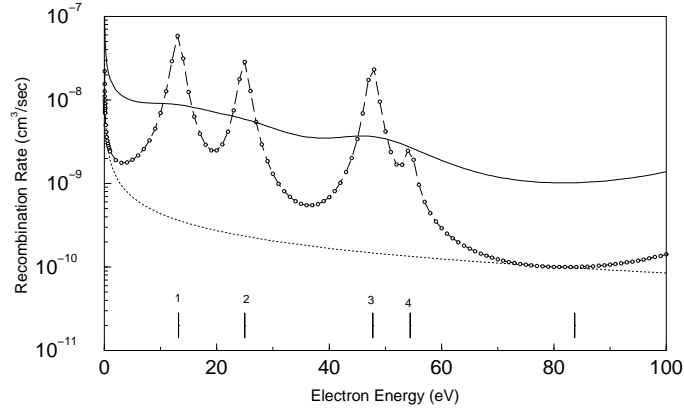}
\caption{Recombination rate in Au$^{52+}$. Dotted line: RR rate, solid
line DR rate using spreading width 0.85 a.u. and open circles conected by
dashed line: DR rate with spreading width 0.75 a.u..The vertical lines
numbered 1 - 4 indicate the position of the doubly excited states
involving inner shell hole.1: 3d$^{3}_{3/2}$3d$^{5}_{5/2}$4d$^{1}_{3/2}$4f$^{1}_{5/2}$,
2 :3d$^{3}_{3/2}$3d$^{5}_{5/2}$4d$^{1}_{3/2}$4f$^{1}_{7/2}$, 3:
 3d$^{3}_{3/2}$3d$^{5}_{5/2}$4d$^{1}_{5/2}$4f$^{1}_{5/2}$ and 4: 3d$^{3}_{3/2}$3d$^{5}_{5/2}$4d$^{1}_{5/2}$4f$^{1}_{7/2}$.}
\label{fig:orb}
\end{figure}

\begin{figure}[h]
\epsfxsize=8cm
\centering\leavevmode\epsfbox{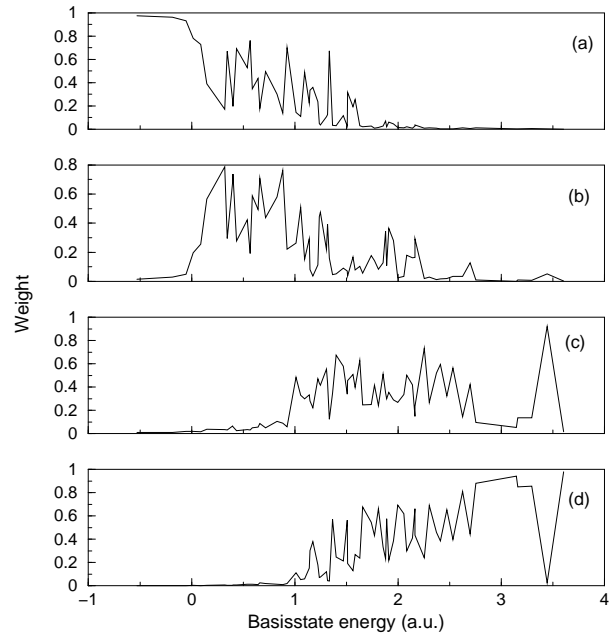}
\caption {Weights of dielectronic states.(a):
3d$^{3}_{3/2}$3d$^{5}_{5/2}$4d$^{1}_{3/2}$4f$^{1}_{5/2}$,
(b):3d$^{3}_{3/2}$3d$^{5}_{5/2}$4d$^{1}_{3/2}$4f$^{1}_{7/2}$, (c):
3d$^{3}_{3/2}$3d$^{5}_{5/2}$4d$^{1}_{5/2}$4f$^{1}_{5/2}$ and (d): 3d$^{3}_{3/2}$3d$^{5}_{5/2}$4d$^{1}_{5/2}$4f$^{1}_{7/2}$.}
\end{figure}

\begin{figure}[h]
\epsfxsize=8cm
\centering\leavevmode\epsfbox{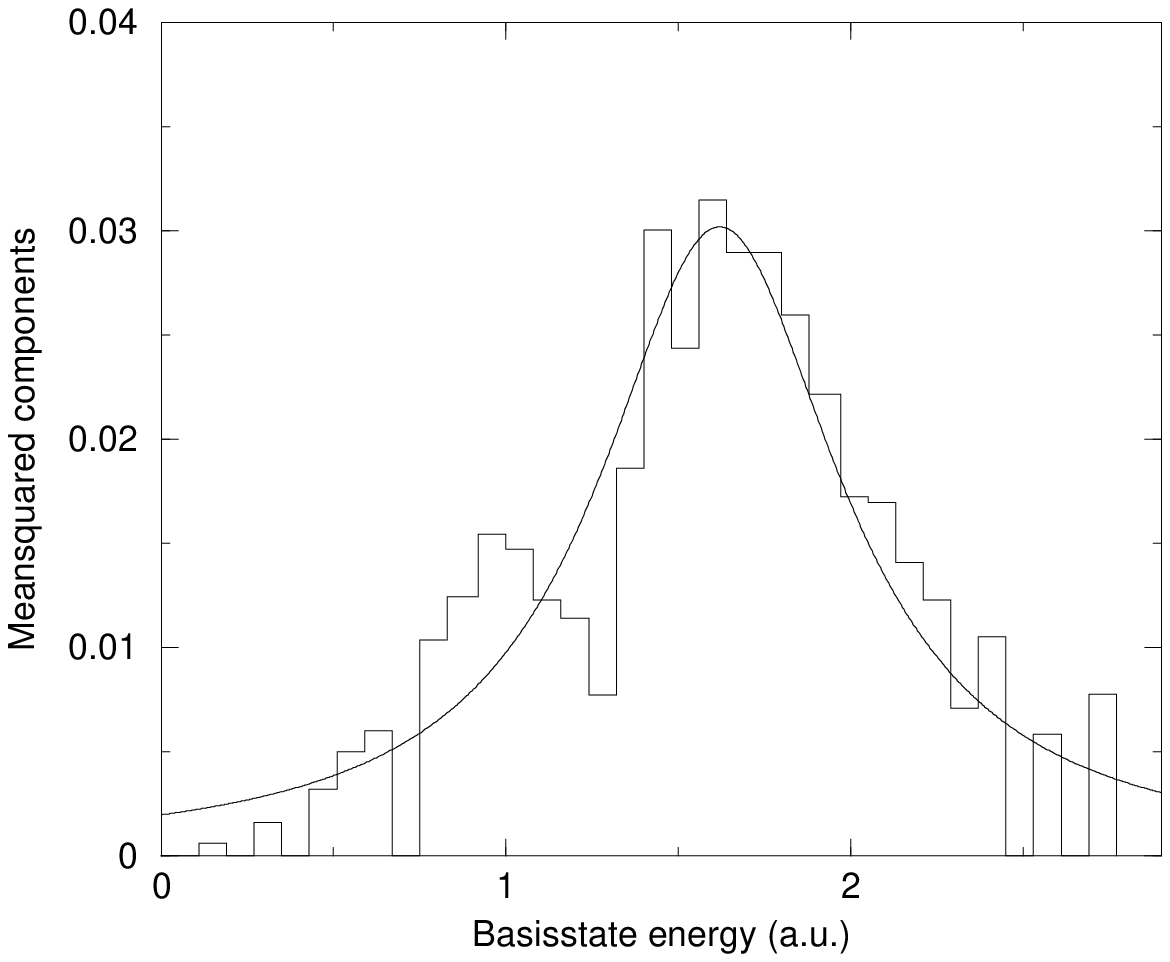}
\caption {Meansquared components avaraged over 11 neighbouring eigen state
components taken from the middle of 69 state CI calculation. The solid
line is the BW fit whic estimates the spreading width is 0.85 a.u..}
\end{figure}

we performed another calculation since we believe that
this value gives an estimate of the real spreading of the doubly excited
states. The results are shown as solid lines in the calculation. The
most striking feature in this curve is that there is  an enhancement of
the DR rate over RR (dotted line). Although the dielectronic states are
positioned far from the ionization threshold, but they show a
substantial mixing with each other. We believe this mixing leads to an
enhancement of DR rate ove RR in this ion. It will be more interesting if more
theoretical calculations as well as experimental measurements are
performed for a complete understanding of this system.

\newpage
\section{Summary and outlook}

In summary, we find the the inner shell excitations paly a major role
in electron recombination with multiply charged complex ions. The inclusion of
the inner shell hole significantly changes the magnuitude of DR rates
throughout the energy range considered. In U$^{28+}$, a similar
observation has been reported by us \cite{Sahoo}. However, we found
in Au$^{50+}$ and Pb$^{53+}$ the dielectronic configurations involving
an inner shell hole (either 3d$^{3/3}$ or 3d$^{5/2}$) are positioned
very close to the ionization threshold and are responssible for
producing higer DR rate over RR in the low electron energies. Also at
higher energies due to the presence of these type of configuration
an enhancement of DR rate is observed. In Au$^{52+}$, we made a
prediction for the rate enhancement since the dielectronic states show a
substantial mixing with each other and associated with a large spreading
width $\sim$ 0.85 a.u.. Certainly we invite other theoretical and experimental
investigations in order to get a clear understanding of DR process for
this system.   
\section{acknowledgements}

We thank Prof. A. M\"uller for providing experimental data in numerical form.

\section{References}




\begin{thebibliography}{99}
\bibitem{comment4} Here we assume that the electron energy is below the target
excitation threshold.
\end{thebibliography}
\end{document}